\documentclass[12pt]{article}
\usepackage{mathrsfs}
\usepackage{amsfonts}
\usepackage{amsmath}
\usepackage{amssymb}
\usepackage{mathptmx}
\usepackage[dvips]{color}
\usepackage{epsfig}
\usepackage{graphicx}
\newcommand{\dslash}{/\!\!\!\!\!\partial}

\newcommand{\tr}{\textrm{tr}}

\begin{document}
\baselineskip=16pt

\pagenumbering{arabic}

\vspace{1.0cm}

\begin{center}
{\Large\sf Generalized Fierz identities and applications to
spin-3/2 particles}
\\[10pt]
\vspace{.5 cm}

{Yi Liao$^a$\footnote{liaoy@nankai.edu.cn}, Ji-Yuan Liu$^b$}

{\it $^a$ School of Physics, Nankai University, Tianjin 300071, China
\\
$^b$ College of Science, Tianjin University of Technology, Tianjin
300384, China}

\vspace{2.0ex}

{\bf Abstract}

\end{center}

We study the most general Fierz identities for a pair of
non-contracted Dirac matrices both in the standard basis and for
chiral spinors. These identities are useful in building independent
effective operators of fermions that involve derivatives or sextuple
and more fields. We apply them to construct independent effective
four-fermion interactions that contain one to three spin-3/2 chiral
fields. These operators could be relevant to physics of higher-spin
baryons and phenomenology of a neutral, massive spin-3/2 particle as
a dark matter candidate.

%
%
%



\newpage

\section{Introduction}
\label{sec:intro}

The Fierz identities are a set of relations that reshuffle a pair
of fermion fields in a product of bilinear forms. The identities
were originally built for a scalar product of two bilinear forms
made of spin-1/2 fields, relating
$\bar\psi_1\Gamma^A\psi_2\bar\psi_3\Gamma_A\psi_4$ to a sum of the
terms $\bar\psi_1\Gamma^B\psi_4\bar\psi_3\Gamma_B\psi_2$
\cite{Fierz, Good:1955em}. Here $\Gamma^A$ refers to a set of
sixteen matrices that are classified as the scalar, pseudoscalar,
vector, axial-vector and tensor, and that together form the
standard basis for $4\times 4$ matrices. This was generalized some
years ago \cite{Nieves:2003in} to the case of the pseudoscalar
products,
$\bar\psi_1\Gamma^A\psi_2\bar\psi_3\tilde\Gamma_A\psi_4$, where
$\tilde\Gamma^A$ stands for the parity-opposite counterpart of
$\Gamma^A$, and the case in which one reshuffles with the help of
charge conjugation a barred and an unbarred spinor (e.g., $\psi_2$
and $\bar\psi_3$) in a scalar or pseudoscalar product.

In practical work we sometimes come across a product
$\bar\psi_1\Gamma^A\psi_2\bar\psi_3\Gamma_B\psi_4$ where the two
indices are not or only partially contracted. Reshuffling a pair of
spinors in this case requires further generalized Fierz identities
for a non-contracted product. Such an identity was first explicitly
written down in Ref. \cite{Nishi:2004st} for the product,
$\bar\psi_1\sigma_{\mu\nu}P_-\psi_2\bar\psi_3\gamma^\nu P_+\psi_4$,
where $P_\pm=(1\pm\gamma_5)/2$ are the projectors for the right- and
left-handed spinors. It proves to be very useful to remove the
redundancy \cite{Girlanda:2008ts} in the leading, dimension-seven,
parity-violating four-nucleon operators in which one Lorentz index
is carried by a derivative \cite{Zhu:2004vw}. A second circumstance
in which such generalized identities are necessary involves
reshuffling spinors in an operator containing six or more spinor
fields. An outstanding example of this is the instanton-induced 't
Hooft six-quark interaction \cite{'tHooft:1986nc} that respects the
chiral $SU(3)$ but breaks the axial $U(1)$ symmetries. Generalized
identities are useful to recast the interaction in a form that would
meet the special needs in the mean-field approach to the
color-superconductor \cite{Steiner:2005jm}. Finally, non-contracted
pairs of gamma matrices appear definitely in less studied
four-fermion operators where spin-3/2 (or higher) fields are present
which carry a Lorentz index. These operators are relevant to hadron
physics that describes effective interactions between the spin-3/2
and spin-1/2 baryons. Rather recently a neutral spin-3/2 particle
has been suggested as a dark matter candidate and its four-fermion
effective interactions were phenomenologically studied
\cite{Kamenik:2011vy,Yu:2011by,Ding:2012sm}.

In this work we conduct a systematic analysis on the non-contracted
Fierz identities. We first do so in the next section in the standard
basis of the gamma matrices. This is followed by section
\ref{sec:chiral} in which chiral spinors are employed. The
identities turn out to be much simpler than in the standard basis,
and their usefulness is demonstrated by a few examples of sextuple
forms of spinors. In section \ref{sec:application} we construct the
quadruple forms involving one to three vector-spinors corresponding
to spin-3/2 particles, and employ the generalized Fierz identities
established in this work to examine their independence. We briefly
summarize in the last section.

\section{Identities in the standard basis}
\label{sec:standard}

The Fierz identities are based on the completeness of a set of 16
matrices for the $4\times 4$ matrices. For the standard basis we
choose the following one:
\begin{eqnarray}
\Gamma^A=1;~\gamma_5;~\gamma^a;~\gamma^a\gamma_5;~\sigma^{a_1a_2}
~(a_1<a_2)
\end{eqnarray}
We work in the convention: $g^{\mu\nu}=\textrm{diag}(+1,-1,-1,-1)$,
$\gamma_5=i\gamma^0\gamma^1\gamma^2\gamma^3$, $\epsilon^{0123}=+1$,
and $\sigma^{\mu\nu}=\frac{i}{2}[\gamma^\mu,\gamma^\nu]$. To
normalize the basis in a simple way, we choose the basis with a
lower index as
\begin{eqnarray}
\Gamma_B=1;~\gamma_5;~\gamma_b; ~\gamma_5\gamma_b;~\sigma_{b_1b_2}
~(b_1<b_2)
\end{eqnarray}
Note that the order of the factors for the axial vector has been
deliberately flipped, so that the orthogonality condition
$\tr(\Gamma^A\Gamma_B)=4\delta^A_B$ holds uniformly.

An arbitrary $4\times 4$ matrix $M$ can be expanded in the basis
as $M=\displaystyle\sum_Ax_A\Gamma^A$, with $4x_A=\tr(M\Gamma_A)$
by the orthogonality condition, i.e.,
$4M=\displaystyle\sum_A\Gamma^A\tr(M\Gamma_A)$. The arbitrariness
of $M$ implies
$\displaystyle\sum_C(\Gamma_C)_{ef}(\Gamma^C)_{ad}=4\delta_{ed}\delta_{af}$,
and multiplying it by $(\Gamma^A)_{fg}$ and summing over $f$ gives
\begin{eqnarray}
\sum_C(\Gamma_C\Gamma^A)_{eg}(\Gamma^C)_{ad}=4(\Gamma^A)_{ag}\delta_{ed}
\end{eqnarray}
Taking the product of the above equality with another that is
obtained from it by the substitutions $egad\to gecb$ and $CA\to DB$,
and summing over $ge$ yields finally
\begin{eqnarray}
4(\Gamma^A)_{ab}(\Gamma^B)_{cd}
=\frac{1}{4}\sum_{CD}\tr(\Gamma_C\Gamma^A\Gamma_D\Gamma^B)(\Gamma^C)_{ad}(\Gamma^D)_{cb}
\end{eqnarray}
which is essentially the most general Fierz identity. One point
deserves special attention. In the sum over $C$ (similarly with $D$)
the six independent components of the tensor are included only once;
to recover the more convenient convention of Lorentz contraction we
include a factor of $1/2$ when $C$ refers to the tensor term. Using
the brief notations for the direct products of matrices, the above
will be denoted as
\begin{eqnarray}
4\Gamma^A\otimes\Gamma^B=\eta_C\eta_D
\frac{1}{4}\tr(\Gamma_C\Gamma^A\Gamma_D\Gamma^B)\Gamma^C\odot\Gamma^D
\end{eqnarray}
where $\eta_C=1/2$ when $C$ refers to the tensor and $\eta_C=1$
otherwise (and similarly for $\eta_D$). In terms of the spin-1/2
field operators the above means
\begin{eqnarray}
-4\bar\psi_1\Gamma^A\psi_2\bar\psi_3\Gamma^B\psi_4=\eta_C\eta_D
\frac{1}{4}\tr(\Gamma_C\Gamma^A\Gamma_D\Gamma^B)
\bar\psi_1\Gamma^C\psi_4\bar\psi_3\Gamma^D\psi_2
\end{eqnarray}
where the minus sign on the left hand side takes care of the
Grassmannian nature of the fields and does not appear when the
identities are applied to the spinor wavefunctions.

We record in what follows the explicit results upon working out
the traces. There are only fifteen independent identities instead
of twenty five; for instance, one can obtain the result for
$\gamma_5\otimes 1$ from that for $1\otimes\gamma_5$ by
interchanging the two matrices in each product on both sides.

The products involving $\Gamma^A=1$ are found to be
\begin{eqnarray}
4[1\otimes 1]&=&[1\odot 1]+[\gamma_5\odot\gamma_5]
+[\gamma^c\odot\gamma_c]-[\gamma^c\gamma_5\odot\gamma_c\gamma_5]
+\frac{1}{2}[\sigma^{c_1c_2}\odot\sigma_{c_1c_2}]%
\label{eq_SS}
\\
4[1\otimes\gamma_5]&=&[1\odot\gamma_5+\leftrightarrow]
+[\gamma_c\gamma_5\odot\gamma^c-\leftrightarrow]
+\frac{1}{2}i[\sigma^{c_1c_2}\odot\tilde\sigma_{c_1c_2}]%
\label{eq_SP}
\end{eqnarray}
\begin{eqnarray}
4[1\otimes\gamma^b]&=&\big[\big(1\odot\gamma^b
-\gamma_c\gamma_5\odot\tilde\sigma^{bc}\big)+\leftrightarrow\big]
+\big[\big(\gamma_5\odot\gamma^b\gamma_5
+i\gamma_c\odot\sigma^{cb}\big)-\leftrightarrow\big]
\\
4[1\otimes\gamma^b\gamma_5]&=& \big[\big(1\odot\gamma^b\gamma_5
-\gamma_c\odot\tilde\sigma^{bc}\big)+\leftrightarrow\big]
+\big[\big(\gamma_5\odot\gamma^b
+i\gamma_c\gamma_5\odot\sigma^{cb}\big)-\leftrightarrow\big]
\\
4[1\otimes\sigma^{b_1b_2}]&=&\big[\big(1\odot\sigma^{b_1b_2}
+i\gamma_5\odot\tilde\sigma^{b_1b_2}
-\epsilon^{b_1b_2cd}\gamma_c\odot\gamma_d\gamma_5\big)+\leftrightarrow\big]
\nonumber\\
&&+i\big[\big(\gamma^{b_2}\odot\gamma^{b_1}
+\gamma^{b_1}\gamma_5\odot\gamma^{b_2}\gamma_5
+\sigma^{b_2d}\odot\sigma^{b_1}_{~~d}\big)-\leftrightarrow\big]
\label{eq_ST}
\end{eqnarray}
where $\leftrightarrow$ means the interchange of the two matrices in
each direct product of the forgoing terms, and the shortcut
$\tilde\sigma^{\mu\nu}\equiv
\frac{1}{2}\epsilon^{\mu\nu\alpha\beta}\sigma_{\alpha\beta}
=-i\sigma^{\mu\nu}\gamma_5$ has been introduced.

The products involving $\Gamma^A=\gamma_5$ are
\begin{eqnarray}
4[\gamma_5\otimes\gamma_5]&=&[1\odot 1]+[\gamma_5\odot\gamma_5]
-[\gamma^c\odot\gamma_c]+[\gamma^c\gamma_5\odot\gamma_c\gamma_5]
+\frac{1}{2}[\sigma^{c_1c_2}\odot\sigma_{c_1c_2}]%
\label{eq_PP}
\\
4[\gamma_5\otimes\gamma^b]&=&\big[\big(1\odot\gamma^b\gamma_5
+\gamma_c\odot\tilde\sigma^{bc}\big)-\leftrightarrow\big]
+\big[\big(\gamma_5\odot\gamma^b
+i\gamma_c\gamma_5\odot\sigma^{bc}\big)+\leftrightarrow\big]
\\
4[\gamma_5\otimes\gamma^b\gamma_5]&=&\big[\big(1\odot\gamma^b
+\gamma_c\gamma_5\odot\tilde\sigma^{bc}\big)-\leftrightarrow\big]
+\big[\big(\gamma_5\odot\gamma^b\gamma_5
+i\gamma_c\odot\sigma^{bc}\big)+\leftrightarrow\big]
\\
4[\gamma_5\otimes\sigma^{b_1b_2}]&=&\big[\big(i1\odot\tilde\sigma^{b_1b_2}
+\gamma_5\odot\sigma^{b_1b_2}
+i(\gamma^{b_2}\odot\gamma^{b_1}\gamma_5-~^{b_1}\leftrightarrow ~^{b_2})\big)%
+\leftrightarrow\big]
\nonumber\\
&&+\epsilon^{b_1b_2cd}\big[-\gamma_c\odot\gamma_d+\gamma_c\gamma_5\odot\gamma_d\gamma_5
+\sigma_{ce}\odot\sigma_{d}^{~~e}\big]
\nonumber\\
&&+\big[\big(\tilde\sigma^{b_1d}\odot\sigma^{b_2}_{~~d}
-~^{b_1}\leftrightarrow ~^{b_2}\big)+\leftrightarrow\big]
\label{eq_PT}
\end{eqnarray}
while the products involving $\Gamma^A=\gamma^a$ are
\begin{eqnarray}
4[\gamma^a\otimes\gamma^b]&=&\big[\big(i1\odot\sigma^{ab}
+\gamma_5\odot\tilde\sigma^{ab}
+i\epsilon^{abcd}\gamma_c\odot\gamma_d\gamma_5\big)-\leftrightarrow\big]
\nonumber
\\
&&+g^{ab}\big[1\odot 1-\gamma_5\odot\gamma_5\big]
+\big[\gamma^a\odot\gamma^b+\gamma^b\odot\gamma^a-g^{ab}\gamma^c\odot\gamma_c\big]
\nonumber
\\
&&
+\big[\gamma^a\gamma_5\odot\gamma^b\gamma_5+\gamma^b\gamma_5\odot\gamma^a\gamma_5
-g^{ab}\gamma^c\gamma_5\odot\gamma_c\gamma_5\big]
\nonumber
\\
&&+\frac{1}{2}\big[g^{ab}\sigma^{c_1c_2}\odot\sigma_{c_1c_2}-2\sigma^{ac}\odot\sigma^b_{~~c}
-2\sigma^{bc}\odot\sigma^a_{~~c}\big]%
\label{eq_VV}
\\
4[\gamma^a\otimes\gamma^b\gamma_5]&=&
g^{ab}\big[\gamma_5\odot 1-\leftrightarrow\big]%
+i\epsilon^{abcd}\big[\gamma_c\odot\gamma_d+\gamma_c\gamma_5\odot\gamma_d\gamma_5\big]
\nonumber\\
&&+\big[(\gamma^a\odot\gamma^b\gamma_5+\gamma^b\odot\gamma^a\gamma_5
-g^{ab}\gamma^c\odot\gamma_c\gamma_5)+\leftrightarrow\big]
\nonumber\\
&&+\big[\big(1\odot\tilde\sigma^{ab}+i\gamma_5\odot\sigma^{ab}\big)
+\leftrightarrow\big]%
+i\epsilon^{abcd}\sigma_{ce}\odot\sigma^e_{~~d}
\nonumber\\
&&+i\big[\sigma^a_{~~c}\odot\tilde\sigma^{bc}-\leftrightarrow\big]
-i\big[\sigma^b_{~~c}\odot\tilde\sigma^{ac}+\leftrightarrow\big]
+\frac{1}{2}ig^{ab}\sigma_{c_1c_2}\odot\tilde\sigma^{c_1c_2}%
\label{eq_VA}
\end{eqnarray}
and
\begin{eqnarray}
&&4[\gamma^a\otimes\sigma^{b_1b_2}]
\nonumber\\
&=&i\big[\big(g^{ab_2}1\odot\gamma^{b_1}-~^{b_1}\leftrightarrow~^{b_2}\big)-\leftrightarrow\big]
+i\big[\big(g^{ab_1}\gamma_5\odot\gamma^{b_2}\gamma_5-~^{b_1}\leftrightarrow~^{b_2}\big)
+\leftrightarrow\big]
\nonumber\\
&&-\epsilon^{ab_1b_2d}\big[1\odot\gamma_d\gamma_5+\leftrightarrow\big]
+\epsilon^{ab_1b_2d}\big[\gamma_5\odot\gamma_d-\leftrightarrow\big]
-i\epsilon^{ab_1b_2d}\big[\gamma^c\gamma_5\odot\sigma_{cd}-\leftrightarrow\big]
\nonumber\\
&&+\big[\big(g^{ab_1}\gamma^c\odot\sigma^{b_2}_{~~c}-g^{ab_2}\gamma^c\odot\sigma^{b_1}_{~~c}
+\gamma^a\odot\sigma^{b_1b_2}+\gamma^{b_1}\odot\sigma^{ab_2}-\gamma^{b_2}\odot\sigma^{ab_1}\big)
+\leftrightarrow\big]
\nonumber\\
&&+i\big[\big(\epsilon^{ab_2cd}\gamma_c\gamma_5\odot\sigma^{b_1}_{~~d}
-~^{b_1}\leftrightarrow~^{b_2}\big)+\leftrightarrow\big]
+i\epsilon^{b_1b_2cd}\big[\gamma_c\gamma_5\odot\sigma^a_{~~d}-\leftrightarrow\big]
\nonumber\\
&&+i\big[\big((g^{ab_2}\gamma_c\gamma_5\odot\tilde\sigma^{b_1c}
+\gamma^{b_2}\gamma_5\odot\tilde\sigma^{ab_1})
-~^{b_1}\leftrightarrow~^{b_2}\big)+\leftrightarrow\big]
\nonumber\\
&&+i\big[\gamma^a\gamma_5\odot\tilde\sigma^{b_1b_2}-\leftrightarrow\big]
\end{eqnarray}

The products involving $\Gamma^A=\gamma^a\gamma_5$ are
\begin{eqnarray}
4[\gamma^a\gamma_5\otimes\gamma^b\gamma_5]&=&g^{ab}\big[-1\odot
1+\gamma_5\odot\gamma_5\big]
-\frac{1}{2}g^{ab}\big[\sigma^{c_1c_2}\odot\sigma_{c_1c_2}\big]
\nonumber\\
&&+\big[\sigma^{ac}\odot\sigma^b_{~~c}+\leftrightarrow\big]
+\big[\gamma^a\odot\gamma^b+\gamma^b\odot\gamma^a-g^{ab}\gamma^c\odot\gamma_c\big]
\nonumber\\
&&+\big[\gamma^a\gamma_5\odot\gamma^b\gamma_5+\gamma^b\gamma_5\odot\gamma^a\gamma_5
-g^{ab}\gamma^c\gamma_5\odot\gamma_c\gamma_5\big]
\nonumber\\
&&-i\big[\big(1\odot\sigma^{ab}-\gamma_5\odot\sigma^{ab}\gamma_5
-\epsilon^{abcd}\gamma_c\odot\gamma_d\gamma_5\big)-\leftrightarrow\big]%
\label{eq_AA}
\end{eqnarray}
and
\begin{eqnarray}
&&4(\gamma^a\gamma_5\otimes\sigma^{b_1b_2})
\nonumber\\
&=&-\epsilon^{ab_1b_2c}\big[1\odot\gamma_c+\leftrightarrow\big]
+i\big[\big(g^{ab_2}1\odot\gamma^{b_1}\gamma_5-~^{b_1}\leftrightarrow~^{b_2}\big)
-\leftrightarrow\big]
\nonumber\\
&&+\epsilon^{ab_1b_2c}\big[\gamma_5\odot\gamma_c\gamma_5-\leftrightarrow\big]
+i\big[\big(g^{ab_1}\gamma_5\odot\gamma^{b_2}-~^{b_1}\leftrightarrow~^{b_2}\big)+
\leftrightarrow\big]
\nonumber\\
&&+i\big[\big(\epsilon^{b_1b_2cd}\gamma_c\odot\sigma^{ad}
-\epsilon^{ab_1b_2d}\gamma^c\odot\sigma_{cd}\big)-\leftrightarrow\big]
+i\big[\big(\epsilon^{ab_2cd}\gamma_c\odot\sigma^{b_1d}
-~^{b_1}\leftrightarrow~^{b_2}\big)+\leftrightarrow\big]
\nonumber\\
&&+\big[\big((g^{ab_2}\gamma_c\odot\sigma^{b_1c}\gamma_5
+\gamma^{b_2}\odot\sigma^{ab_1}\gamma_5)
-~^{b_1}\leftrightarrow~^{b_2}\big)+\leftrightarrow\big]
+\big[\gamma^a\odot\sigma^{b_1b_2}\gamma_5-\leftrightarrow\big]
\nonumber\\
&&+\big[\big((g^{ab_1}\gamma_c\gamma_5\odot\sigma^{b_2c}
+\gamma^{b_1}\gamma_5\odot\sigma^{ab_2})-~^{b_1}\leftrightarrow~^{b_2}\big)
+\leftrightarrow\big]
+\big[\gamma^a\gamma_5\odot\sigma^{b_1b_2}+\leftrightarrow\big]
\end{eqnarray}

Finally, the expression for the product
$\sigma^{a_1a_2}\otimes\sigma^{b_1b_2}$ is very lengthy:
\begin{eqnarray}
&&4(\sigma^{a_1a_2}\otimes\sigma^{b_1b_2})
\nonumber\\
&=&(g^{a_1b_1}g^{a_2b_2}-~^{b_1}\leftrightarrow~^{b_2})\big[1\odot 1
+\gamma_5\odot\gamma_5+\gamma^c\odot\gamma_c
-\gamma^c\gamma_5\odot\gamma_c\gamma_5
+\frac{1}{2}\sigma^{c_1c_2}\odot\sigma_{c_1c_2}\big]
\nonumber\\
&&
+i\epsilon^{a_1a_2b_1b_2}\big[1\odot\gamma_5+\leftrightarrow\big]%
-i\epsilon^{a_1a_2b_1b_2}\big[\gamma_c\odot\gamma^c\gamma_5-\leftrightarrow\big]
\nonumber\\
&&+i\big[\big((g^{a_1b_1}1\odot\sigma^{a_2b_2}+g^{a_2b_2}1\odot\sigma^{a_1b_1})
-~^{b_1}\leftrightarrow~^{b_2}\big)-\leftrightarrow\big]
\nonumber\\
&&+\big[\big(\epsilon^{a_1a_2b_1d}\gamma_5\odot\sigma^{b_2}_{~~d}
-~^{b_1}\leftrightarrow~^{b_2}\big)+\leftrightarrow\big]
+\big[\big(\epsilon^{a_2b_1b_2d}\gamma_5\odot\sigma^{a_1}_{~~d}
-~^{a_1}\leftrightarrow~^{a_2}\big)-\leftrightarrow\big]
\nonumber\\
&&+\big[\big((g^{a_1b_1}\gamma_5\odot\tilde\sigma^{a_2b_2}
+g^{a_2b_2}\gamma_5\odot\tilde\sigma^{a_1b_1})
-~^{b_1}\leftrightarrow~^{b_2}\big)+\leftrightarrow\big]
\nonumber\\
&&+\big[\big((g^{a_2b_1}\gamma^{b_2}\odot\gamma^{a_1}
+g^{a_1b_2}\gamma^{b_1}\odot\gamma^{a_2})
-~^{b_1}\leftrightarrow~^{b_2}\big)+\leftrightarrow\big]
\nonumber\\
&&-\big[\big((g^{a_2b_1}\gamma^{b_2}\gamma_5\odot\gamma^{a_1}\gamma_5
+g^{a_1b_2}\gamma^{b_1}\gamma_5\odot\gamma^{a_2}\gamma_5)
-~^{b_1}\leftrightarrow~^{b_2}\big)+\leftrightarrow\big]
\nonumber\\
&&+i\big[\big(\epsilon^{a_1a_2b_2c}(\gamma_c\odot\gamma^{b_1}\gamma_5
-\gamma^{b_1}\odot\gamma_c\gamma_5)-~^{b_1}\leftrightarrow~^{b_2}\big)+\leftrightarrow\big]
\nonumber\\
&&+i\big[\big(\epsilon^{a_1b_1b_2c}(\gamma_c\odot\gamma^{a_2}\gamma_5
+\gamma^{a_2}\odot\gamma_c\gamma_5)-~^{a_1}\leftrightarrow~^{a_2}\big)-\leftrightarrow\big]
\nonumber\\
&&+i\big[\big((g^{a_1b_1}\epsilon^{a_2b_2cd}+g^{a_2b_2}\epsilon^{a_1b_1cd})
-~^{b_1}\leftrightarrow~^{b_2}\big)
\gamma_c\odot\gamma_d\gamma_5+\leftrightarrow\big]
\nonumber\\
&& +\big[\big(\sigma^{a_1a_2}\odot\sigma^{b_1b_2}
+\sigma^{a_1b_1}\odot\sigma^{a_2b_2}
+\sigma^{b_1a_2}\odot\sigma^{a_1b_2}\big)+\leftrightarrow\big]
\nonumber\\
&&+\big[\big((g^{a_1b_1}\sigma^{ca_2}\odot\sigma^{b_2}_{~~c}
+g^{a_2b_2}\sigma^{cb_1}\odot\sigma^{a_1}_{~~c})-~^{a_1}\leftrightarrow~^{a_2}\big)
+\leftrightarrow\big]%
\label{eq_TT}
\end{eqnarray}

The above identities are the most general ones for $4\times 4$
matrices, and can reproduce as special cases the fully or partially
Lorentz-contracted ones reported in the literature. For instance,
the standard Lorentz-scalar Fierz identities are given by eqs
(\ref{eq_SS},\ref{eq_PP}) and eqs
(\ref{eq_VV},\ref{eq_AA},\ref{eq_TT}) upon contracting all Lorentz
indices with the signature tensor, while the generalized
Lorentz-pseudoscalar identities listed in \cite{Nieves:2003in} are
given by eq (\ref{eq_SP}), eq (\ref{eq_VA}) upon contraction by
$g_{ab}$, and eq (\ref{eq_TT}) upon contraction by
$\epsilon_{a_1a_2b_1b_2}$.

\section{Identities for chiral spinors}
\label{sec:chiral}

The Fierz identities are usually used to rearrange the fields in
effective operators. Since the fermion fields in the standard model
are chiral, those operators are naturally given in terms of chiral
fields. It is thus desirable in this case to have the generalized
identities projected onto various combinations of the two
chiralities. There are two equivalent approaches to work them out.
One can either choose a (partially) chiral basis, e.g.,
$P_\mp,~\gamma^aP_\mp,~\sigma^{ab}$, and proceed as in the last
section, or obtain the identities by directly applying the
projectors $P_\mp$ to the results recorded above. We have computed
in both approaches and arrived at identical results. It is very nice
to find that employing chiral fields significantly simplifies the
identities.

We first introduce some notations. We use the shortcuts
$\psi_{1\mp}=P_\mp\psi_1$ and
$\bar\psi_{1\mp}=(\psi_{1\mp})^\dagger\gamma_0=\bar\psi_1P_\pm$.
Here $\psi_1$ is a fermion field but can be spin-1/2 or spin-3/2 (or
even higher). In the latter case $\psi_1$ implicitly carries a
Lorentz index. We call
$\bar\psi_1\Gamma^A\psi_{2\mp}\bar\psi_3\Gamma^B\psi_{4\mp}$ as
chirality-diagonal and
$\bar\psi_1\Gamma^A\psi_{2\mp}\bar\psi_3\Gamma^B\psi_{4\pm}$ as
chirality-flipped. For a given pair of $\Gamma^{A,B}$, the
chiralities of $\bar\psi_1$ and $\bar\psi_3$ are accordingly fixed
and will be carried over to the Fierz-rearranged terms. This is
natural and sufficient in practice since one cannot change the
chirality of a field by applying an algebraic relation. In other
words, the identities to be presented below apply to the spinor
fields of definite chiralities.

We start with the simplest identity in eq (\ref{eq_SS}). We
project it out by attaching $P_\mp$ from right to the two matrices
(identity in this case) in the product on the left-hand side, so
that we will obtain a Fierz identity for
$\bar\psi_{1\pm}\psi_{2\mp}\bar\psi_{3\pm}\psi_{4\mp}$, where the
chirality of $\psi_1$ ($\psi_3$) is singled out by that of
$\psi_2$ ($\psi_4$). One can see that the first two terms on the
right-hand side become equal while the third and fourth terms are
killed due to the mismatch in chirality. The result in terms of
field operators is
\begin{eqnarray}
-4\bar\psi_{1\pm}\psi_{2\mp}\bar\psi_{3\pm}\psi_{4\mp}%
=2\bar\psi_{1\pm}\psi_{4\mp}\bar\psi_{3\pm}\psi_{2\mp}
+\frac{1}{2}\bar\psi_{1\pm}\sigma^{c_1c_2}\psi_{4\mp}\bar\psi_{3\pm}\sigma_{c_1c_2}\psi_{2\mp}
\end{eqnarray}
We will write the above as an algebraic relation:
\begin{eqnarray*}
4[P_\mp\otimes P_\mp]\sim 2[P_\mp\odot P_\mp]
+\frac{1}{2}[\sigma^{c_1c_2}P_\mp\odot\sigma_{c_1c_2}P_\mp]
\end{eqnarray*}
where the similarity symbol becomes equality when the spinor
wavefunctions (field operators) are attached (with a minus sign on
the left). In a similar manner, we can obtain the
chirality-flipped identity by projecting eq (\ref{eq_SS}) with
$P_\mp\cdots P_\pm$ on its left side. Note that the projectors are
interchanged on its right side, becoming $P_\pm\cdots P_\mp$, so
that the third and fourth terms become equal while all other three
are killed:
\begin{eqnarray*}
4[P_\mp\otimes P_\pm]\sim 2[\gamma^cP_\pm\odot\gamma_cP_\mp]
\end{eqnarray*}
The above two relations can also be obtained by starting from eq
(\ref{eq_SP}), which serves as a consistency check of the result
in the standard basis. In so doing for the chirality-diagonal
case, the relation (\ref{eq_basic}) in Appendix is employed. The
generalized Fierz identities to be listed in the following all
pass similar consistency checks.

For brevity, we first present the results and then make some brief
comments on the derivation. The generalized Fierz identities are,
in the chirality-diagonal case,
\begin{eqnarray}
4[P_\mp\otimes P_\mp]&\sim&2[P_\mp\odot P_\mp]
+\frac{1}{2}[\sigma^{c_1c_2}P_\mp\odot\sigma_{c_1c_2}P_\mp]
\label{eq_SSdiag}\\
4[P_\mp\otimes\gamma^bP_\mp]&\sim&2[P_\mp\odot\gamma^bP_\mp]
+2i[\sigma^{bc}P_\mp\odot\gamma_cP_\mp]
\label{eq_SVdiag}\\
4[P_\mp\otimes\sigma^{b_1b_2}P_\mp]
&\sim&2[P_\mp\odot\sigma^{b_1b_2}P_\mp+\leftrightarrow]
+i[\sigma^{b_2d}P_\mp\odot\sigma^{b_1}_{~~d}P_\mp-\leftrightarrow]%
\label{eq_STdiag}\\
4[\gamma^aP_\mp\otimes\gamma^bP_\mp]%
&\sim&2[\gamma^aP_\mp\odot\gamma^bP_\mp+\gamma^bP_\mp\odot\gamma^aP_\mp
-g^{ab}\gamma^cP_\mp\odot\gamma_cP_\mp]
\nonumber\\
&&\mp 2i\epsilon^{abcd}\gamma_cP_\mp\odot\gamma_dP_\mp
\label{eq_VVdiag}\\
4[\gamma^aP_\mp\otimes\sigma^{b_1b_2}P_\mp]
&\sim&2i\big[g^{ab_1}\gamma^{b_2}P_\mp\odot P_\mp-~^{b_1}\leftrightarrow~^{b_2}\big]%
\nonumber\\
&&\pm 2\epsilon^{ab_1b_2d}\gamma_dP_\mp\odot P_\mp
+2\gamma^aP_\mp\odot\sigma^{b_1b_2}P_\mp
\nonumber\\
&&+2\big[\big(g^{ab_1}\gamma_cP_\mp\odot\sigma^{b_2c}P_\mp
+\gamma^{b_1}P_\mp\odot\sigma^{ab_2}P_\mp\big)
-~^{b_1}\leftrightarrow~^{b_2}\big]
\label{eq_VTdiag}\\
4[\sigma^{a_1a_2}P_\mp\otimes\sigma^{b_1b_2}P_\mp]&\sim&
\frac{1}{2}\big(g^{a_1b_1}g^{a_2b_2}-~^{b_1}\leftrightarrow~^{b_2}\big)
\big[4P_\mp\odot P_\mp-\sigma^{c_1c_2}P_\mp\odot\sigma_{c_1c_2}P_\mp\big]%
\nonumber\\
&&+\big[\sigma^{a_1a_2}P_\mp\odot\sigma^{b_1b_2}P_\mp+\leftrightarrow\big]
\mp 2i\epsilon^{a_1a_2b_1b_2}P_\mp\odot P_\mp%
\nonumber\\
&&+2i\big[P_\mp\odot\big((g^{a_1b_1}\sigma^{a_2b_2}+g^{a_2b_2}\sigma^{a_1b_1})
-~^{b_1}\leftrightarrow~^{b_2}\big)P_\mp-\leftrightarrow\big]
\nonumber\\
&&+\big[\big(\sigma^{a_1b_1}P_\mp\odot\sigma^{a_2b_2}P_\mp
-~^{a_1}\leftrightarrow~^{a_2}\big)+\leftrightarrow\big]
\label{eq_TTdiag}
\end{eqnarray}
and in the chirality-flipped case,
\begin{eqnarray}
4[P_\mp\otimes P_\pm]&\sim& 2[\gamma^cP_\pm\odot\gamma_cP_\mp]
\label{eq_SSflip}\\
4[P_\mp\otimes\gamma^bP_\pm]&\sim&2[\gamma^bP_\pm\odot P_\mp]
+2i[\gamma_cP_\pm\odot\sigma^{cb}P_\mp]
\label{eq_SVflip}\\
4[P_\mp\otimes\sigma^{b_1b_2}P_\pm]
&\sim&2i[\gamma^{b_2}P_\pm\odot\gamma^{b_1}P_\mp-~^{b_1}\leftrightarrow~^{b_2}]%
\pm 2\epsilon^{b_1b_2cd}[\gamma_cP_\pm\odot\gamma_dP_\mp]
\label{eq_STflip}\\
4[\gamma^aP_\mp\otimes\gamma^bP_\pm]%
&\sim&2g^{ab}[P_\pm\odot P_\mp]
-\big[\sigma^{ac}P_\pm\odot\sigma^b_{~~c}P_\mp+~^a\leftrightarrow~^b\big]
\nonumber\\
&& +2i\big[P_\pm\odot\sigma^{ab}P_\mp-\leftrightarrow\big]
\label{eq_VVflip}\\
4[\gamma^aP_\mp\otimes\sigma^{b_1b_2}P_\pm]
&\sim&2i\big[g^{ab_2}P_\pm\odot\gamma^{b_1}P_\mp-~^{b_1}\leftrightarrow~^{b_2}\big]%
\nonumber\\
&&\pm 2\epsilon^{ab_1b_2d}P_\pm\odot\gamma_dP_\mp
+2\sigma^{b_1b_2}P_\pm\odot\gamma^aP_\mp
\nonumber\\
&&+2\big[\big(g^{ab_1}\sigma^{b_2}_{~~c}P_\pm\odot\gamma^cP_\mp
+\sigma^{ab_2}P_\pm\odot\gamma^{b_1}P_\mp\big)
-~^{b_1}\leftrightarrow~^{b_2}\big]
\label{eq_VTflip}\\
4[\sigma^{a_1a_2}P_\mp\otimes\sigma^{b_1b_2}P_\pm]
&\sim&2\big[\big((g^{a_2b_1}\gamma^{b_2}P_\pm\odot\gamma^{a_1}P_\mp
+g^{a_1b_2}\gamma^{b_1}P_\pm\odot\gamma^{a_2}P_\mp)
\nonumber\\
&&-~^{b_1}\leftrightarrow~^{b_2} \big)+\leftrightarrow\big]
+2\big(g^{a_1b_1}g^{a_2b_2}-g^{a_1b_2}g^{a_2b_1}\big)\gamma^cP_\pm\odot\gamma_cP_\mp
\nonumber\\
&&\mp 2i\big(\epsilon^{a_1a_2b_2c}g^{b_1d}-\epsilon^{a_1a_2b_1c}g^{b_2d}%
\nonumber\\
&& +\epsilon^{a_1b_1b_2d}g^{a_2c}-\epsilon^{a_2b_1b_2d}g^{a_1c}\big)
\gamma_cP_\pm\odot\gamma_dP_\mp%
\label{eq_TTflip}
\end{eqnarray}
Note that the interchange of matrices $\leftrightarrow$ does not
act on the chiral projectors; for instance, the second term in the
last square brackets of eq (\ref{eq_VVflip}) reads,
$-\sigma^{ab}P_\pm\odot P_\mp$.

In checking eq (\ref{eq_VVflip}) obtained from eq (\ref{eq_VV})
against the one from eq (\ref{eq_VA}) or eq (\ref{eq_AA}), eqs
(\ref{eq_rel1},\ref{eq_rel2},\ref{eq_rel3}) in Appendix are used. In
deriving eq (\ref{eq_TTdiag}), eqs (\ref{eq_rel7},\ref{eq_rel2p})
are applied, while  eqs (\ref{eq_rel5},\ref{eq_rel6}) are employed
to cast eq (\ref{eq_TTflip}) in the displayed form. Eq
(\ref{eq_rel5}) is also useful in simplifying eqs
(\ref{eq_VTdiag},\ref{eq_VTflip}), while eq (\ref{eq_rel7}) is used
to recast eq (\ref{eq_VTdiag}) in a form similar to eq
(\ref{eq_VTflip}). Finally, eq (\ref{eq_rel4}) is used in checking
eq (\ref{eq_STdiag}) obtained from eq (\ref{eq_ST}) against that
from eq (\ref{eq_PT}).

We illustrate the above results by a few examples of sextuple
forms in spin-1/2 fields. The first one concerns the operators
$A_\mp=\bar\psi_{1\pm}\psi_{2\mp}\bar\psi_{3\pm}\psi_{4\mp}\bar\psi_{5\pm}\psi_{6\mp}$.
They have the same Lorentz structure as the instanton-induced
six-quark interaction.  We will not try to transform them to a
form that would be relevant to the study of
color-superconductivity \cite{Steiner:2005jm}, which involves
rearrangements of the barred and unbarred fields with charge
conjugation as well as nontrivial structures in flavor and color
spaces. Instead, we are content with transforming them to a form
with $\psi_{2,4,6}$ replaced by $\psi_{4,6,2}$. This can be
accomplished by twice applications of eqs
(\ref{eq_SSdiag}-\ref{eq_TTdiag}). Although the procedures are not
unique, the end result must be the same. For instance, we may
first apply eq (\ref{eq_SSdiag}) to
$\bar\psi_{1\pm}\psi_{2\mp}\bar\psi_{3\pm}\psi_{4\mp}$ and then
eqs (\ref{eq_SSdiag},\ref{eq_STdiag}) to
$\bar\psi_{3\pm}\Gamma^A\psi_{2\mp}\bar\psi_{5\pm}\psi_{6\mp}$.
The Grassmannian minus signs are cancelled in the final result:
\begin{eqnarray}
16A_\mp&=&
4\bar\psi_{1\pm}\psi_{4\mp}\bar\psi_{3\pm}\psi_{6\mp}\bar\psi_{5\pm}\psi_{2\mp}%
+i\bar\psi_{1\pm}\sigma_{dc}\psi_{4\mp}\bar\psi_{3\pm}\sigma^c_{~e}\psi_{6\mp}
\bar\psi_{5\pm}\sigma^{ed}\psi_{2\mp}
\nonumber\\
&&+\bar\psi_{1\pm}\psi_{4\mp}\bar\psi_{3\pm}\sigma^{cd}\psi_{6\mp}
\bar\psi_{5\pm}\sigma_{cd}\psi_{2\mp}%
+\bar\psi_{1\pm}\sigma_{cd}\psi_{4\mp}\bar\psi_{3\pm}\sigma^{cd}\psi_{6\mp}
\bar\psi_{5\pm}\psi_{2\mp}%
\nonumber\\
&&+\bar\psi_{1\pm}\sigma_{cd}\psi_{4\mp}\bar\psi_{3\pm}\psi_{6\mp}
\bar\psi_{5\pm}\sigma^{cd}\psi_{2\mp}
\end{eqnarray}
Consider next the operators with a contracted vector,
$B_\mp=\bar\psi_{1\mp}\gamma^a\psi_{2\mp}\bar\psi_{3\mp}\gamma_a\psi_{4\mp}
\bar\psi_{5\pm}\psi_{6\mp}$, which can be Fierz-rewritten as
\begin{eqnarray}
2B_\mp&=&-\bar\psi_{1\mp}\gamma^a\psi_{4\mp}\big[\bar\psi_{3\mp}
\gamma_a\psi_{6\mp}\bar\psi_{5\pm}\psi_{2\mp}+i\bar\psi_{3\mp}
\gamma^b\psi_{6\mp}\bar\psi_{5\pm}\sigma_{ab}\psi_{2\mp}\big]
\end{eqnarray}
Finally, compare a pair of operators that differ by chirality in one
bilinear,
\begin{eqnarray}
C_\mp&=&\bar\psi_{1\mp}\gamma^a\psi_{2\mp}\bar\psi_{3\mp}\gamma^b\psi_{4\mp}
\bar\psi_{5\pm}\sigma_{ab}\psi_{6\mp}
\\
D_\mp&=&\bar\psi_{1\mp}\gamma^a\psi_{2\mp}\bar\psi_{3\pm}\gamma^b\psi_{4\pm}
\bar\psi_{5\pm}\sigma_{ab}\psi_{6\mp}
\end{eqnarray}
It is easier to first apply eqs (\ref{eq_VVdiag},\ref{eq_VVflip})
and keep terms antisymmetric in $a$ and $b$, and then use eqs
(\ref{eq_VTdiag},\ref{eq_STflip},\ref{eq_TTflip}) to obtain
\begin{eqnarray}
4C_\mp&=&\bar\psi_{1\mp}\gamma_cP_\mp\psi_{4\mp}\big[
-6i\bar\psi_{3\mp}\gamma^{c}\psi_{6\mp}\bar\psi_{5\pm}\psi_{2\mp}%
-2\bar\psi_{3\mp}\gamma_d\psi_{6\mp}\bar\psi_{5\pm}\sigma^{cd}\psi_{2\mp}\big]
\\
4D_\mp&=&i\bar\psi_{1\mp}\psi_{4\pm}
\big[12\bar\psi_{3\pm}\psi_{6\mp}\bar\psi_{5\pm}\psi_{2\mp}
-4\bar\psi_{3\pm}\sigma_{ab}\psi_{6\mp}\bar\psi_{5\pm}\sigma^{ab}\psi_{2\mp}\big]
\nonumber\\
&&-\bar\psi_{1\mp}\sigma^{ab}\psi_{4\pm}\bar\psi_{3\pm}\sigma_{bd}\psi_{6\mp}
\bar\psi_{5\pm}\sigma^d_{~~a}\psi_{2\mp}
\end{eqnarray}
where eqs (\ref{eq_basic},\ref{eq_rel1}) have been used.

\section{Applications to spin-3/2 particles}
\label{sec:application}

The quantum field of a spin-3/2 particle can be described by the
Rarita-Schwinger (RS) field \cite{Rarita:1941mf}, $\Psi_\mu$. It
is a vector-spinor that has the mixed transformation properties of
a Dirac field and a vector field under Lorentz transformations. A
free field of mass $M$ satisfies the Dirac equation,
$(i\dslash-M)\Psi_\mu=0$, with the constraint
\begin{eqnarray}
\gamma^\mu\Psi_\mu=0%
\label{eq_constraint}
\end{eqnarray}
from which follows the relation $\partial^\mu\Psi_\mu=0$. For our
purpose of constructing dimension-six four-fermion effective
operators without a derivative, only the constraint
(\ref{eq_constraint}) and its conjugate $\bar\Psi_\mu\gamma^\mu=0$
are relevant. In the following, we will construct those operators
involving up to three factors of the RS fields that may be
relevant to the phenomenology of the standard model particles, and
when necessary verify their independence using the Fierz
identities established in the last section.

We start with the operators containing a single RS field,
$\Psi_\mu$, and three spin-1/2 chiral fields,
$\psi_{1\mp},~\psi_{2\mp},~\psi_{3\mp}$, i.e., of the types:
\begin{eqnarray}
X^\textrm{d}_\mp&=&\bar\psi_1\Gamma^AP_\mp\psi_2\bar\psi_3\Gamma^BP_\mp\Psi_\mu
\\
X^\textrm{f}_\pm&=&\bar\psi_1\Gamma^AP_\mp\psi_2\bar\psi_3\Gamma^BP_\pm\Psi_\mu
\end{eqnarray}
where all Lorentz indices are to be properly contracted. Allocating
the basis matrices to both $\Gamma^A$ and $\Gamma^B$ with an odd
number of Lorentz indices yields the possibilities,
$1\otimes\gamma^b,~\gamma^a\otimes 1,
~\gamma^a\otimes\sigma^{b_1b_2},~\sigma^{a_1a_2}\otimes\gamma^b$,
where $\gamma_5$ is excluded due to the presence of $P_\mp$. Now we
contract the above indices with $\Psi_\mu$. It is not necessary
either to employ the Levi-Civita tensor since, e.g.,
$\epsilon^{a_1a_2cd}\sigma_{a_1a_2}P_\mp=2\tilde\sigma^{cd}P_\mp=\pm
2i\sigma^{cd}P_\mp$ using eq (\ref{eq_basic}), which does not yield
a new form. The first form is killed by eq (\ref{eq_constraint}).
The third one is removed as redundant, since
$\sigma_{a\mu}P_\mp\Psi^\mu=i(\gamma_a\gamma_\mu-g_{a\mu})P_\mp
\Psi^\mu=-iP_\mp\Psi_a$, which is the second form in the list. The
list is thus shortened to $\gamma^\mu\otimes
1,~\sigma^{\mu\nu}\otimes\gamma_\nu$, corresponding to the operators
\begin{eqnarray}
X^\textrm{d}_\mp&=&\bar\psi_1\gamma^\mu P_\mp\psi_2\bar\psi_3P_\mp\Psi_\mu,
~\bar\psi_1\sigma^{\mu\nu}P_\mp\psi_2\bar\psi_3\gamma_\nu P_\mp\Psi_\mu
\\
X^\textrm{f}_\pm&=&\bar\psi_1\gamma^\mu P_\mp\psi_2\bar\psi_3P_\pm\Psi_\mu,
~\bar\psi_1\sigma^{\mu\nu}P_\mp\psi_2\bar\psi_3\gamma_\nu P_\pm\Psi_\mu
\end{eqnarray}
With the help of the identities in section \ref{sec:chiral} it can
be verified that they are complete and independent, i.e., that one
cannot get lesser or more operators by interchanging the roles of
$\psi_2$ and $\Psi_\mu$ (or equivalently $\psi_1$ and $\psi_3$).
Consider for instance the operators obtained from the second pair
in $X_\mp^\textrm{d}$ by $\bar\psi_1\leftrightarrow\bar\psi_3$.
Using eq (\ref{eq_VTdiag}) upon contracting a pair of indices
\begin{eqnarray*}
&&4[\gamma_\nu P_\mp\otimes\sigma^{\mu\nu}P_\mp]
\\
&\sim&%
2i(-3)\gamma^\mu P_\mp\odot P_\mp +2\gamma_\nu
P_\mp\odot\sigma^{\mu\nu}P_\mp\odot \pm
2i\epsilon^{\nu\mu\alpha\beta}\gamma_\alpha
P_\mp\odot\sigma_{\nu\beta}P_\mp
\\
&=&%
2i(-3)\gamma^\mu P_\mp\odot P_\mp -2\gamma_\nu
P_\mp\odot\sigma^{\mu\nu}P_\mp
\end{eqnarray*}
where eq (\ref{eq_basic}) is used in the last term, we have upon
applying the constraint (\ref{eq_constraint})
\begin{eqnarray}
-4\bar\psi_1\gamma_\nu P_\mp\Psi_\mu\bar\psi_3\sigma^{\mu\nu}P_\mp\psi_2%
=-8i\bar\psi_1\gamma^\mu P_\mp\psi_2\bar\psi_3P_\mp\Psi_\mu
\end{eqnarray}
which is indeed the first pair in $X_\mp^\textrm{d}$.

The operators involving two spin-1/2 and two spin-3/2 fields are
classified into two classes. In the first class we can write in
either the format
\begin{eqnarray}
\textrm{(Ia)}&&Y_\mp^\textrm{(Ia)d}=\bar\Psi_{1\mu}\Gamma^AP_\mp\Psi_{2\nu}
\bar\psi_1\Gamma^BP_\mp\psi_2
\\
&&Y_\pm^\textrm{(Ia)f}=\bar\Psi_{1\mu}\Gamma^AP_\mp\Psi_{2\nu}\bar\psi_1\Gamma^BP_\pm\psi_2
\end{eqnarray}
or its Fierz-transformed one
\begin{eqnarray}
\textrm{(Ib)}&&
Y_\mp^\textrm{(Ib)d}=\bar\Psi_{1\mu}\Gamma^AP_\mp\psi_2\bar\psi_1\Gamma^BP_\mp\Psi_{2\nu}
\\
&&Y_\pm^\textrm{(Ib)f}=\bar\Psi_{1\mu}\Gamma^AP_\mp\psi_2\bar\psi_1\Gamma^BP_\pm\Psi_{2\nu}
\end{eqnarray}
We keep both for the purpose of redundancy check. In the second
class, we have
\begin{eqnarray}
\textrm{(II)}&&
Y_\mp^\textrm{(II)d}=\bar\psi_1\Gamma^AP_\mp\Psi_{1\mu}\bar\psi_2\Gamma^BP_\mp\Psi_{2\nu}
\\
&&Y_\pm^\textrm{(II)f}=\bar\psi_1\Gamma^AP_\mp\Psi_{1\mu}\bar\psi_2\Gamma^BP_\pm\Psi_{2\nu}
\end{eqnarray}
Note that the operators in class (II) violate any additive quantum
number carried by the $\psi$ fields, e.g., the lepton or baryon
number, or by the $\Psi_\mu$ fields. The pair of $\Gamma^{A,B}$
should have an even number of indices, $1\otimes
1;~\gamma^a\otimes\gamma^b;
~1\otimes\sigma^{b_1b_2},~\sigma^{a_1a_2}\otimes 1;
~\sigma^{a_1a_2}\otimes\sigma^{b_1b_2}$, to be contracted with the
field operators.

Consider first class (Ia). The possible forms without involving a
$\sigma$ are
\begin{eqnarray*}
&&g^{\mu\nu}P_\mp\otimes P_\mp;%
~g^{\mu\nu}\gamma^aP_\mp\otimes\gamma_aP_\mp,%
~\epsilon^{\mu\nu ab}\gamma_aP_\mp\otimes\gamma_bP_\mp
\\
&&g^{\mu\nu}P_\mp\otimes P_\pm;%
~g^{\mu\nu}\gamma^aP_\mp\otimes\gamma_aP_\pm,%
~\epsilon^{\mu\nu ab}\gamma_aP_\mp\otimes\gamma_bP_\pm
\end{eqnarray*}
Due to eq (\ref{eq_basic}), the possible forms containing one
$\sigma$ are restricted to be
\begin{eqnarray*}
&&P_\mp\otimes\sigma_{\mu\nu}P_\mp
\\
&&P_\mp\otimes\sigma_{\mu\nu}P_\pm
\end{eqnarray*}
With two $\sigma$'s, the completely self-contracted ones are
\begin{eqnarray*}
g^{\mu\nu}\sigma^{a_1a_2}P_\mp\otimes\sigma_{a_1a_2}P_\mp
\end{eqnarray*}
whose chirality-flipped counterparts,
$g^{\mu\nu}\sigma^{a_1a_2}P_\mp\otimes\sigma_{a_1a_2}P_\pm$,
vanish by eq (\ref{eq_rel1}). The terms with once self-contracted
$\sigma$'s are either reducible to those with one $\sigma$ in the
chirality-diagonal case, e.g,
$\sigma^\mu_{~~c}P_\mp\otimes\sigma^{\nu c}P_\mp$ and
$\epsilon^{\mu\nu ab}\sigma_a^{~c}P_\mp\otimes\sigma_{bc}P_\mp$
with the help of eqs (\ref{eq_constraint},\ref{eq_rel4}), or
simply vanish by eqs (\ref{eq_rel1},\ref{eq_rel2}) in the
chirality-flipped case. It thus appears that
$Y_\mp^\textrm{(Ia)d}$ have five forms and $Y_\pm^\textrm{(Ia)f}$
have four. A similar analysis shows that class (Ib) has the same
number of forms:
\begin{eqnarray*}
&&g^{\mu\nu}P_\mp\otimes P_\mp;%
~g^{\mu\nu}\gamma^aP_\mp\otimes\gamma_aP_\mp,%
~\gamma^\nu P_\mp\otimes\gamma^\mu P_\mp,%
~\epsilon^{\mu\nu ab}\gamma_aP_\mp\otimes\gamma_bP_\mp;
~g^{\mu\nu}\sigma^{ab}P_\mp\otimes\sigma_{ab}P_\mp
\\
&&g^{\mu\nu}P_\mp\otimes P_\pm;%
~g^{\mu\nu}\gamma^aP_\mp\otimes\gamma_aP_\pm,%
~\gamma^\nu P_\mp\otimes\gamma^\mu P_\pm,%
~\epsilon^{\mu\nu ab}\gamma_aP_\mp\otimes\gamma_bP_\pm
\end{eqnarray*}

But the lists for class (Ia) and (Ib) are actually redundant. To
see this, we can start from either (Ia) or (Ib). Starting from
(Ia) (i.e., $\otimes$ for (Ia) and $\odot$ for (Ib)) and using eqs
(\ref{eq_VVdiag},\ref{eq_VVflip},\ref{eq_basic}), we have
\begin{eqnarray}
4[\gamma_aP_\mp\otimes\gamma_bP_\mp]\epsilon^{\mu\nu ab}%
&\sim&\pm 4i[\gamma^\mu P_\mp\odot\gamma^\nu P_\mp
-\gamma^\nu P_\mp\odot\gamma^\mu P_\mp]%
\label{eq_Iabdiag}\\
4[\gamma_aP_\mp\otimes\gamma_bP_\pm]\epsilon^{\mu\nu ab}%
&\sim&\pm 4i[P_\pm\odot\sigma^{\mu\nu}P_\mp+\sigma^{\mu\nu}P_\pm\odot P_\mp]%
\label{eq_Iabflip}
\end{eqnarray}
The first term in eq (\ref{eq_Iabdiag}) is killed by eq
(\ref{eq_constraint}) while the second term remains in the list of
(Ib). Both terms in eq (\ref{eq_Iabflip}) reduce to
$g^{\mu\nu}P_\pm\odot P_\mp$ by eq (\ref{eq_constraint}). The
above identities also apply when starting from (Ib) (now $\otimes$
for (Ib) and $\odot$ for (Ia)). Then, both terms in eq
(\ref{eq_Iabdiag}) are killed, while the first term in eq
(\ref{eq_Iabflip}) remains and the second term reduces to
$g^{\mu\nu}P_\pm\odot P_\mp$. The conclusion from this is that
$\gamma_aP_\mp\otimes\gamma_bP_\mp\epsilon^{\mu\nu ab}$ and
$\gamma_aP_\mp\otimes\gamma_bP_\pm\epsilon^{\mu\nu ab}$ can be
simultaneously removed from the list in both (Ia) and (Ib). This
result conforms to the list of fourteen independent operators in
\cite{Ding:2012sm} in the standard basis. Finally, the analysis
and result for (Ib) apply to class (II) as well.

We finally come to the operators involving three spin-$3/2$ and
one spin-1/2 fields:
\begin{eqnarray}
Z^\textrm{d}_\mp&=&\bar\Psi_{1\alpha}\Gamma^AP_\mp
\Psi_{2\beta}\bar\psi\Gamma^BP_\mp\Psi_{3\gamma}
\\
Z^\textrm{f}_\pm&=&\bar\Psi_{1\alpha}\Gamma^AP_\mp\Psi_{2\beta}\bar\psi\Gamma^BP_\pm
\Psi_{3\gamma}
\end{eqnarray}
The pair of $\Gamma^{A,B}$ with an odd number of indices includes
$\gamma^a\otimes 1,~1\otimes\gamma^b;~
\gamma^a\otimes\sigma^{b_1b_2},~\sigma^{a_1a_2}\otimes\gamma^b$,
and thus the possible forms for both $Z^\textrm{d,f}_\mp$ are
\begin{eqnarray}
&&g^{\alpha\beta}\gamma^\gamma\otimes 1,
~\epsilon^{\alpha\beta\gamma a}\gamma_a\otimes 1%
\label{eq_line1}
\\
&&g^{\alpha\gamma}1\otimes\gamma^\beta,~g^{\beta\gamma}1\otimes\gamma^\alpha,
~\epsilon^{\alpha\beta\gamma b}1\otimes\gamma_b%
\label{eq_line2}
\\
&&\gamma^\gamma\otimes\sigma^{\alpha\beta},
~g^{\beta\gamma}\gamma_a\otimes\sigma^{a\alpha},
~g^{\alpha\gamma}\gamma_a\otimes\sigma^{a\beta},
\label{eq_line3}
\\
&&\epsilon^{\alpha\beta\gamma b}\gamma^a\otimes\sigma_{ab},
~\epsilon^{\alpha\gamma ab}\gamma_a\otimes\sigma^\beta_{~~b},%
~\epsilon^{\beta\gamma ab}\gamma_a\otimes\sigma^\alpha_{~~b}%
\label{eq_line4}
\\
&&g^{\alpha\beta}\sigma^{\gamma b}\otimes\gamma_b,
~\epsilon^{\alpha\beta\gamma a}\sigma_{ab}\otimes\gamma^b,
~\epsilon^{\alpha\beta ab}\sigma^\gamma_{~~a}\otimes\gamma_b
\label{eq_line5}
\end{eqnarray}
The last term in eqs (\ref{eq_line1},\ref{eq_line2}) and the last
two terms in eq (\ref{eq_line4}) can be removed as redundant using
eqs (\ref{eq_basic2},\ref{eq_constraint}). For the first term in
eq (\ref{eq_line4}) and the second term in eq (\ref{eq_line5}),
using antisymmetry in $a,~b$ and eq (\ref{eq_rel5}), we have
\begin{eqnarray*}
\epsilon^{\alpha\beta\gamma a}\gamma^b-\epsilon^{\alpha\beta\gamma b}\gamma^a%
=-\epsilon^{\beta\gamma ab}\gamma^\alpha%
-\epsilon^{\gamma ab\alpha}\gamma^\beta%
-\epsilon^{ab\alpha\beta}\gamma^\gamma%
\end{eqnarray*}
which will transform $\sigma_{ab}$ into
$\tilde\sigma^{\beta\gamma}$ etc and thus can be dropped. Finally,
the last term in eq (\ref{eq_line5}) can be recast using eq
(\ref{eq_rel7}) as, for both chirality-diagonal and -flipped
cases,
\begin{eqnarray*}
\epsilon^{\alpha\beta ab}\sigma^\gamma_{~~a}P_\mp\otimes\gamma_b%
=\pm i (g^{\beta\gamma}\sigma^{\alpha b}
-g^{\alpha\gamma}\sigma^{\beta b}
-g^{b\gamma}\sigma^{\alpha\beta})P_\mp\otimes\gamma_b
\end{eqnarray*}
where the first two terms are reducible to those already covered
while the third vanishes by eq (\ref{eq_constraint}). In summary,
the list has been reduced for both $Z^\textrm{d,f}_\mp$ to
\begin{eqnarray}
&&g^{\alpha\beta}\gamma^\gamma\otimes 1,~
g^{\alpha\gamma}1\otimes\gamma^\beta,~g^{\beta\gamma}1\otimes\gamma^\alpha
\nonumber\\
&&\gamma^\gamma\otimes\sigma^{\alpha\beta},~
g^{\beta\gamma}\gamma_a\otimes\sigma^{a\alpha},
g^{\alpha\gamma}\gamma_a\otimes\sigma^{a\beta},~
g^{\alpha\beta}\sigma^{\gamma b}\otimes\gamma_b
\end{eqnarray}
which are obviously independent as are explicitly verified by the
generalized identities.

\section{Summary}
\label{sec:summary}

We have studied the general Fierz identities for all possible direct
products of non-contracted Dirac matrices both in the standard basis
and for chiral fields. The identities are helpful in rearranging
fields in effective four-fermion operators involving derivatives and
in sextuple or higher operators. They can also be applied to
effective interactions involving higher-spin fermions. We have
illustrated this by considering all dimension-six four-fermion
operators that involve one to three spin-3/2 fields, and used the
identities to check their independency. These operators could be
relevant to the phenomenological study of baryons with a higher spin
and of the recent interest in massive spin-3/2 particles as a
possible candidate for dark matter.

\vspace{0.5cm}
\noindent %
{\bf Acknowledgement}

This work was supported in part by the grant NSFC-11025525 and by
the Fundamental Research Funds for the Central Universities
No.65030021.

\vspace{0.5cm}
\noindent %
{\Large\bf Appendix: some useful relations}
\vspace{0.5cm}

We list below some algebraic relations that were employed in
deriving and simplifying the generalized Fierz identities. The most
often used is the basic relation
\begin{eqnarray}
\tilde\sigma^{\mu\nu}P_\mp=\pm i\sigma^{\mu\nu}P_\mp
\label{eq_basic}
\end{eqnarray}
where
$\tilde\sigma^{\mu\nu}=(1/2)\epsilon^{\mu\nu\alpha\beta}\sigma_{\alpha\beta}$.
In four-dimensions, we have
\begin{eqnarray}
p^a\epsilon^{bcde}+p^b\epsilon^{cdea}+p^c\epsilon^{deab}
+p^d\epsilon^{eabc}+p^e\epsilon^{abcd}=0%
\label{eq_rel5}
\end{eqnarray}
where $p$ is an arbitrary four-component quantity, e.g., the gamma
matrix, and
\begin{eqnarray}
&&2\big(g^{a_1b_1}\epsilon^{a_2b_2cd}-g^{a_1b_2}\epsilon^{a_2b_1cd}
-g^{a_2b_1}\epsilon^{a_1b_2cd}+g^{a_2b_2}\epsilon^{a_1b_1cd}\big)
\nonumber\\
&=&+\big[(g^{b_1c}\epsilon^{a_1a_2b_2d}-g^{b_1d}\epsilon^{a_1a_2b_2c})
-(g^{b_2c}\epsilon^{a_1a_2b_1d}-g^{b_2d}\epsilon^{a_1a_2b_1c})\big]
\nonumber\\
&&-\big[(g^{a_1c}\epsilon^{b_1b_2a_2d}-g^{a_1d}\epsilon^{b_1b_2a_2c})
-(g^{a_2c}\epsilon^{b_1b_2a_1d}-g^{a_2d}\epsilon^{b_1b_2a_1c})\big]
\label{eq_rel6}
\end{eqnarray}
which has been constructed by considering all possible six-index
constant tensors made of one signature and one Levi-Civita tensor
with specific symmetries in the indices. Using eq (\ref{eq_basic})
and once-contracted two $\epsilon$'s,
$\epsilon^{a_1a_2b_1d}\epsilon_{dc_1c_2b_2}$, one derives
\begin{eqnarray}
\mp\epsilon^{a_1a_2b_1d}\sigma^{b_2}_{~~d}P_\mp=
i(g^{a_2b_2}\sigma^{a_1b_1}-g^{a_1b_2}\sigma^{a_2b_1}-g^{b_1b_2}\sigma^{a_1a_2})P_\mp
\label{eq_rel7}
\end{eqnarray}
Finally, using the well-known relation
\begin{eqnarray*}
\gamma^\alpha\gamma^\beta\gamma^\gamma=
(g^{\alpha\beta}\gamma^\gamma+g^{\beta\gamma}\gamma^\alpha-g^{\gamma\alpha}\gamma^\beta)
-i\epsilon_\mu^{~~\alpha\beta\gamma}\gamma^\mu\gamma_5
\end{eqnarray*}
the following result is obtained
\begin{eqnarray}
\epsilon^{\alpha\beta\gamma a}\gamma_aP_\mp=%
\mp i\big(\gamma^\alpha\gamma^\gamma\gamma^\beta
-g^{\alpha\gamma}\gamma^\beta-g^{\gamma\beta}\gamma^\alpha
+g^{\alpha\beta}\gamma^\gamma\big)P_\mp%
\label{eq_basic2}
\end{eqnarray}
which was used in section \ref{sec:application} to remove redundant
operators.

Now we list some relations employed in simplifying operators and
Fierz identities. The chirality-flipped ones are
\begin{eqnarray}
&&\sigma^{ab}P_\pm\odot\sigma_{ab}P_\mp=0%
\label{eq_rel1}
\\
&&\sigma^{ac}P_\pm\odot\sigma^b_{~~c}P_\mp=\sigma^{bc}P_\pm\odot\sigma^a_{~~c}P_\mp
\label{eq_rel2}
\\
&&\epsilon^{abcd}\sigma_{ce}P_\pm\odot\sigma^e_{~~d}P_\mp=0
\label{eq_rel3}
\end{eqnarray}
Eq (\ref{eq_rel1}) is derived using eq (\ref{eq_basic}):
\begin{eqnarray*}
\sigma^{\mu\nu}P_\pm\odot\sigma_{\mu\nu}P_\mp&=&%
\mp i\sigma^{\mu\nu}P_\pm\odot\tilde\sigma_{\mu\nu}P_\mp%
=\mp i\frac{1}{2}\epsilon_{\mu\nu\alpha\beta}\sigma^{\mu\nu}P_\pm\odot\sigma^{\alpha\beta}P_\mp%
\\
&=&\mp i\tilde\sigma_{\alpha\beta}P_\pm\odot\sigma^{\alpha\beta}P_\mp%
=-\sigma^{\mu\nu}P_\pm\odot\sigma_{\mu\nu}P_\mp=0
\end{eqnarray*}
Eq (\ref{eq_rel2}) is obtained by expressing both $\sigma$'s in
terms of $\tilde\sigma$ using eq (\ref{eq_basic}), contracting the
two $\epsilon$'s, and applying eq (\ref{eq_rel1}), while eq
(\ref{eq_rel3}) is a result of symmetry in eq (\ref{eq_rel2}).
Similar manipulations in the chirality-diagonal case yield
\begin{eqnarray}
&&\sigma^{ac}P_\mp\odot\sigma^b_{~~c}P_\mp
=\frac{1}{2}g^{ab}\sigma^{cd}P_\mp\odot\sigma_{cd}P_\mp
-\sigma^{bc}P_\mp\odot\sigma^a_{~~c}P_\mp%
\label{eq_rel2p}
\\
&&\mp i\epsilon^{abcd}\sigma_{ce}P_\mp\odot\sigma^e_{~~d}P_\mp
=\sigma^b_{~~d}P_\mp\odot\sigma^{ad}P_\mp-~^a\leftrightarrow^b
\label{eq_rel4}
\end{eqnarray}

\noindent %

\end{document}